\documentclass[superscriptaddress,reprint,showpacs,showkeys,preprintnumbers,amsmath,amssymb,citeautoscript,aps,prb]{revtex4-1}
\usepackage[T1]{fontenc}
\usepackage{graphicx}
\usepackage{color}
\usepackage{epstopdf}

\begin{document}

\preprint{Shekhar et al.}

\title{Large linear magnetoresistance and weak anti-localization in Y(Lu)PtBi topological insulators}

\author{Chandra Shekhar}
\email{cgsbond@gmail.com}
\affiliation{Max Planck Institute for Chemical Physics of Solids, 01187 Dresden, Germany}
\author{Erik Kampert}
\affiliation{Helmholtz Zentrum Dresden Rossendorf, Hochfeld Magnetlab Dresden HLD, 01328 Dresden, Germany}
\author{Tobias F\"{o}rster}
\affiliation{Helmholtz Zentrum Dresden Rossendorf, Hochfeld Magnetlab Dresden HLD, 01328 Dresden, Germany}
\author{Binghai Yan}
\affiliation{Max Planck Institute for Chemical Physics of Solids, 01187 Dresden, Germany}
\author{Ajaya K. Nayak}
\affiliation{Max Planck Institute for Chemical Physics of Solids, 01187 Dresden, Germany}
\author{Michael Nicklas}
\affiliation{Max Planck Institute for Chemical Physics of Solids, 01187 Dresden, Germany}
\author{Claudia Felser}
\email{felser@cpfs.mpg.de}
\affiliation{Max Planck Institute for Chemical Physics of Solids, 01187 Dresden, Germany}

\date{\today}

\begin{abstract}

Topological insulators are new kind of Dirac materials that possess bulk insulating and surface conducting behavior. They are basically known for their unique electronic properties of their surface. Here we present the magneto-transport properties of our high quality single crystalline Y(Lu)PtBi Heusler topological insulators, which belong to a group of noncentrosymmetric superconductor with T$_c=0.8$~K. Both the compounds show semi-metallic behavior with low charge carrier of 6 $\times 10^{18}$ ~cm$^{-3}$ for YPtBi and 8 $\times 10^{19}$~cm$^{-3}$ for LuPtBi at 2~K. Magneto-conductivity measurements in the tilted field indicate that the charge carriers transport through quantum interference. This provide a direct evidence of weak anti-localization below 50~K, giving a strong signature of surface transport. Most importantly, these compounds show very high unsaturated linear magnetoresistance of approximately 2000\% at 200~K in a magnetic field of 60~T.  

\end{abstract}



\maketitle

\begin{center}
\textbf{I. INTRODUCTION}
\end{center}
Topological insulators (TIs) have attracted much attentions because of their unique electronic structure, which comprise of the metallic states on their edges or surfaces.~\cite{MN12,ZA10} As a result of strong spin-orbit coupling, these surface states have special characteristics of locking of spin and momentum, which are protected by time-reversal symmetry. Weak anti-localization (WAL) is one of the hallmarks of the topological surface states, in which conductivity increases due to quantum interference between self-intersecting conduction paths. WAL is destroyed if time reversal symmetry is broken, i.e. a small perpendicular magnetic field is sufficient to cause a clear drop in a conductivity of topological insulators.~\cite{CS14,ZL12,LB12,ML12} Among the well known TIs, half Heusler compounds are promising candidates showing topological properties.\cite{SC10,HL10,DX10,XMZ11,CL11} These compounds (TT'M) usually consist of transition or rare earth metals (T,T') with one main group element (M) in a 1:1:1 stoichiometry. They possess a MgAgAs type face centered cubic (space group $F\:\overline{4}3m$) crystal structure,\cite{TG11} where T, T' and M atoms occupy Wyckoff position 4b, 4c and 4a, respectively. In the case that these compounds contain rare-earth elements, which have strongly correlated $f$ electrons, they show extraordinary physical properties, such a linear dispersion relation of bands along with a {\it zero band} gap having high mobility~\cite{CS12,SO11b,SO11}, half-metallicity~\cite{RAG83}, semiconducting~\cite{KG07}, giant magnetoresistive~\cite{JP00}, heavy fermion~\cite{DK99} and superconducting properties, and various other properties that can be seen in the review.\cite{TG11} Superconductivity in Heusler materials is dominated by unconventional Cooper pairing known as mixed parity, that arises due to antisymmetric spin-orbit coupling, which source is the non-centrosymmetric crystal structure. Moreover, these materials have fascinating electronic properties and a potential application in quantum information processing, magnetoelectric devices, spintronics, and could be used in the realization of Majorana fermions.

A positive, extraordinarily high, linear magnetoresistivity (MR) in nonmagnetic compounds has collected much attraction.~\cite{RX97,TT98,MK09,HGJ10,XW12,CS12} Extensive efforts have been made towards the identification of the origin and involved mechanisms as well as towards material optimization for eventual technological applications, such as high-density data storage or magnetic sensors and actuators. A variety of materials shows a linear MR over wide temperature (mK$<$T$<$400~K) and magnetic field (few mT $< $B$ <$ 60~T) ranges, such as elemental metals,~\cite{KL98,PK29} intermetallic compounds,~\cite{SLB98,MA08} Ag$_{2+\delta}$X (X =Se, Te),~\cite{MK09,ML02,AH01,RX97} InSb,~\cite{JH07} carbon\cite{PMV11} graphene,~\cite{ALF10} graphite,~\cite{SVM05} GaAs-MnAs,~\cite{HGJ10} and BaFe$_2$As$_2$.~\cite{KKH11} For materials with zero gap and a linear dispersion, quantum MR has been proposed as the origin of the linear MR, which is expected to be temperature-independent.~\cite{AAA98} Typical examples of quantum MR observation are in graphene \cite{ALF10}, Ag$_{2+\delta}$X~\cite{SL12}, and Bi$_2$Te$_3$~\cite{XW12}, which have a zero gap, linear Dirac dispersion. On the other hand, a classical MR model has been proposed by Parish et al., where linear MR is expected to be governed by mobility, for materials as GaAs-MnAs composites,~\cite{HGJ10}, Heusler TIs~\cite{CS12, CS12a}, and Ag$_{2+\delta}$Te~\cite{ML02}. In order to shed more light on the origin of linear MR in non-magnetic compounds, we present our MR measurements of Y(Lu)PtBi TIs.


\begin{center}
\textbf{II. EXPERIMENTAL DETAILS}
\end{center}

Single crystals of Y(Lu)PtBi were grown by the solution growth method, whereas Bi acts as a flux. Stoichiometric quantities of freshly polished pieces of elements Y or Lu, Pt and Bi of purity ~$>$99.99\% in the atomic ratio of 0.5:0.5:10 for Y and 0.3:0.3:10 for Lu were put in a tantalum crucible and sealed in a dry quartz ampoule under 3~mbar partial argon pressure. The filled ampoules were heated at a rate of 100~K/h up to 1473~K, followed by 12 hours of soaking. For crystal growth, the temperature was slowly reduced by 2~K/h to 873~K and the surplus of Bi flux was removed by decanting the ampoule at 873~K. Using this method, we obtained 4-5 mm regular triangular shaped crystals, with a preferred growth in the (111)-direction. The general crystal growth procedure was followed from literature.~\cite{PCC92,PCC91} The composition and crystal structure were checked by energy dispersive X-ray analysis and Laue X-ray diffraction, respectively. The lattice parameters of the cubic structure are 6.651~{\AA} for YPtBi and 6.574~{\AA} for LuPtBi, which are consistent with previous reports.~\cite{NPB11,TVB12,FFT13,HL10,WAS10} Energy dispersive x-ray spectrometry yields atomic percentages 34.6 : 32.5 : 32.9 $\pm$ 3.0 \% for Y : Pt : Bi and 33.6 : 33.0 : 33.4 $\pm$ 3.0 \% for Lu : Pt : Bi confirming the stoichiometric ratios of the chemical compositions. Transport measurements were performed using a Quantum Design Physical Property Measurement System (PPMS) with a standard low-frequency lock-in technique in the temperature range from 0.3 to 300~K. Four-probe resistivity and five-probe Hall coefficient measurements were carried out by a conventional AC bridge technique in the Ohmic limit.


\begin{center}
\textbf{III. RESULTS AND DISCUSSION}
\end{center}

\begin{figure}[htb]
\centering
\includegraphics[width=8.5cm]{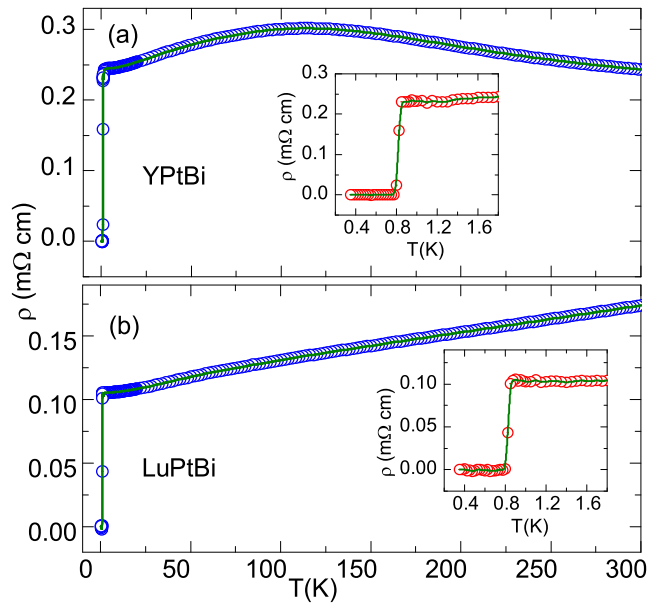}
\caption{(Color online) Temperature dependence resistivity of (a) YPtBi and (b) LuPtBi. The insets show the zoom out view near T${_c}$.}
\label{fig:RT}
\end{figure}

The temperature dependence of the resistivity of a material reflects the scattering of charge carriers with phonons and impurities at high and low temperatures, respectively. Figure ~\ref{fig:RT} shows the temperature dependence of the electrical resistivity $\rho(T)$, of YPtBi and LuPtBi. The observed $\rho$ value of 0.25~m$\Omega$~cm for YPtBi and 0.17~m$\Omega$~cm for LuPtBi at 300~K are comparable with other reports.~\cite{NPB11,TVB12,FFT13} The resistivity of YPtBi increases with decreasing temperature, goes through a broad maximum at T$_{max}$=125~K, whereafter it decreases with decreasing temperature. The resistivity below T$_{max}$ is dominated by WAL, where conductivity is expected to increase. This effect is more pronounced in the magneto-conductivity, which is the main topic of this article. The resistivity of LuPtBi decreases linearly with decreasing temperature, no broad transition is observed. Interestingly, at very low temperatures $\rho$ of both compounds decreases sharply and shows a superconducting onset temperature at 0.85 K for YPtBi and 0.86 K for LuPtBi. $\rho$ becomes zero at 0.8~K for the both compounds, which indicate superconducting transition temperature, T$_c$, and are clearly visible in the insets of Fig.~\ref{fig:RT}. This finding of the T$_c$ is not new in these compounds ~\cite{NPB11,TVB12,FFT13} but recently discovered their quantum phenomenon of topological insulators~\cite{SC10,CL11,HL10,WAS10} has triggered to re-investigate the transport properties.

\begin{figure}[htb]
\centering
 \includegraphics [width=8.5cm]{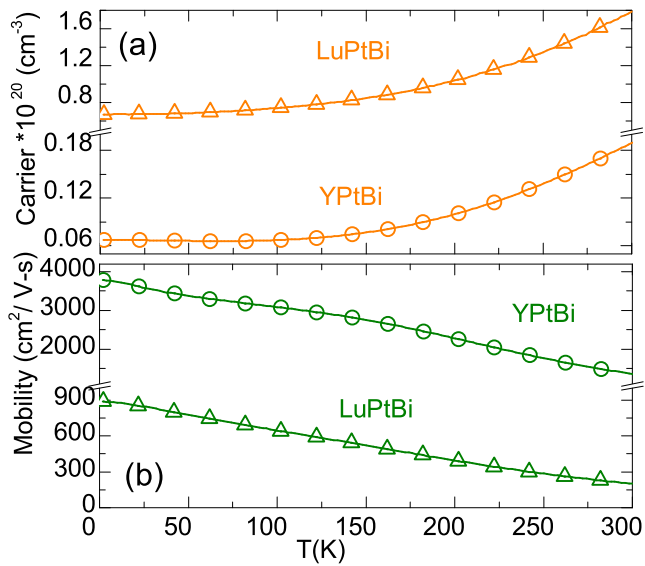}
\caption{(Color online) Temperature dependence of (a) carrier density  and (b) Hall mobility of YPtBi and LuPtBi.}
\label{fig:Hall}
\end{figure}

The experimental investigation of the Hall effect of a compound is a instrumental in deriving informations on the most important kinetic parameters, such as charge carrier concentration and mobility. Therefore, we performed the Hall-effect measurements in temperature sweep as well as field sweep modes of our well characterized single crystals. Both the compounds show positive Hall coefficient $R_H(T)$ through out the temperatures range from 2--300~K, suggesting that holes are the predominant charge carriers. $R_H(T)$ is linear with field in whole temperatures range that indicates the involvement of only one type of charge carrier in transport properties, and nonlinear Hall resistivity shows the involvement of more than one type of carriers.~\cite{DTM96} Thus, the single carrier band Drude model, $n_H(T)=1/n_e R_H(T)$ for carrier density and $\mu_H(T)= R_H(T)/\rho(T)$ for mobility, were applied and their estimated values are shown in Fig.~\ref{fig:Hall}. The values of n$_H$ range from $6.8\times10^{18}$ to $1.9\times10^{19}$~cm$^{-3}$ for YPtBi and $6.7\times10^{19}$ to $1.7\times10^{20}$~cm$^{-3}$ for LuPtBi in the temperature 2--300~K, which are equivalent to other reports~\cite{NPB11,TVB12,FFT13} It can be seen that the carriers density increases as increase temperature showing  semi-metals or low gaped semiconductors like behavior. It is believed that the RPtBi compounds may be either gapless semiconducting or semimetallic, depending on the choice of the rare earth element R.~\cite{SO11,PCC91} This gapless semiconducting is favored by the lighter rare earth atoms while the semimetallic is favored by the heavier ones and these depend on the strength of spin-orbital coupling.~\cite{SC10,WAS10,TO01}

 
The mobility plays a central role for charge transport in a material, which is critically related to the device efficiency. High mobility value reveals high carrier transport, and consequently increases the efficiency of various devices. Both the compounds exhibit high value of the mobility 3790~cm$^2$V$^{-1}$s$^{-1}$ for YPtBi and 887 ~cm$^2$V$^{-1}$s$^{-1}$  for LuPtBi at 2~K. Taking into account of materials properties, therefore, the observed high values of mobility mainly originate from gaplessness of bulk~\cite{CS12a,CS12,SO11,TT98} and linear dispersion of bands in which the charge carriers possess very low effective mass resulting high mobility.\cite{JHC08} These Heusler compounds have both the properties and hence show high mobility. This high mobility adds positively in their topological property, which are important, not only as a physical phenomena but also for further device applications such as image magnetic monopoles, neutral Majorana fermions and giant magneto--optical effects.

\begin{figure}[htb]
\centering
\includegraphics[width=8.5cm]{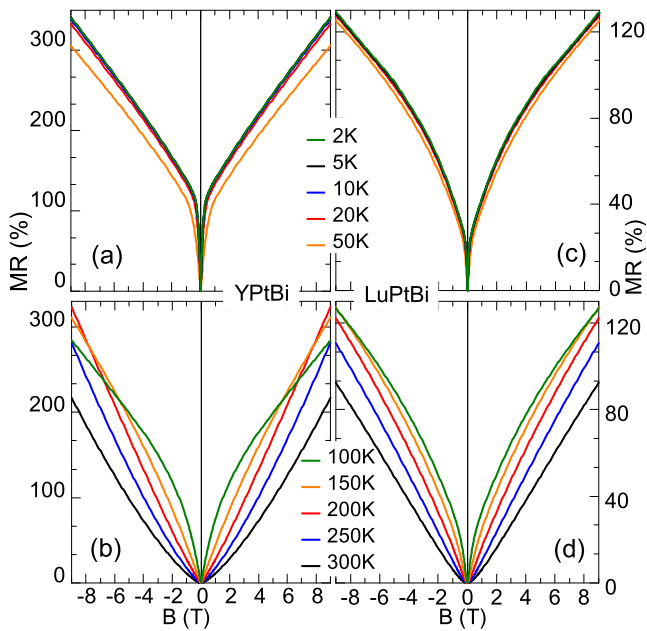}
\caption{(Color online) Isothermal transverse magnetoresistance (field $\bot$ current) of YPtBi and LuPtBi at (a) (c) 2, 5, 10, 20, and 50~K, (b) (d) 100 to 300~K in the step of 50~K.}
\label{fig:MR_T}
\end{figure}

\begin{figure}[htb]
\centering
\includegraphics[width=8.5cm]{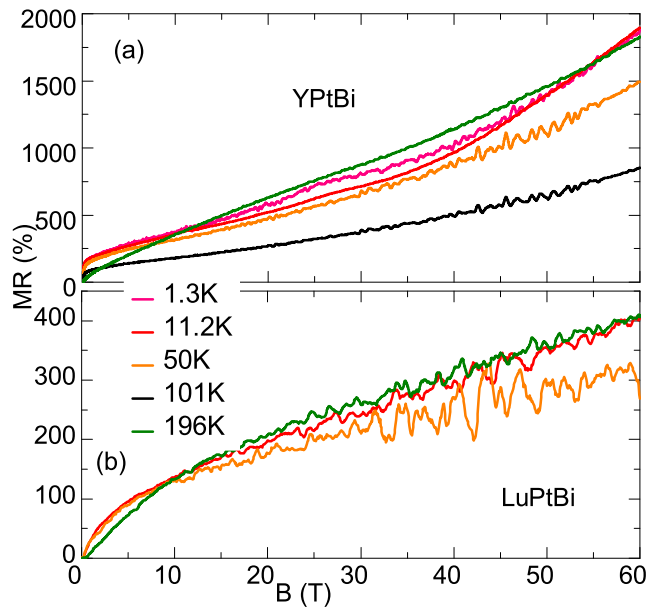}
\caption{(Color online) Isothermal transverse magnetoresistance (field $\bot$ current) of YPtBi and LuPtBi at  different temperature up to 60~T.}
\label{fig:MR_60T}
\end{figure}

For technological applications, the MR is important and its linear response over a large field range makes the materials attractive as magnetic field sensors.~\cite{AH01} The MR is defined as the changes in resistivity with fields, as ${\rm MR}(B)=\rho(B)/\rho(0)-1$, where $B$ is the applied magnetic field. The field dependence transverse MR (B $\bot$ applied current) of YPtBi and LuPtBi at selected temperatures is shown in Fig.~\ref{fig:MR_T}. The followings are the remarkable features of the observed  MR. The overall patterns are positive that show systematic variations with field, which is nearly temperature independent below 20~K. It first sharply increases below a certain threshold of field , and then transforms into a linear rising with increasing magnetic field without showing any sign of saturation. This threshold point mainly depends on temperature and vanishes above 100~K and we will discuss the origin of this behavior separately. The recorded values of MR in 9~T are 200\% at 300~K and 340\% at 2~K for YPtBi, which completly differ in polycrystlline meterilas that MR show only 2--30\%.~\cite{CS13} That means scattering of carriers by frequetly found impurities is not a source of this MR at least in this type of materials.  Since, linear behavior as well as high values of MR of single crystals incomparision to polycrystalline draw a attension to investigate, whether linear behavior will retain or not in ultra high magnetic field. Therefore, we measured resistance in high magnetic pulsed field up to 60~T, which is shown in Fig.~\ref{fig:MR_60T}. The MR at 60~T also show almost linear rising with increasing field without showing any sign of saturation and its value at 200~K is 2000\% in 60~T for YPtBi. These characteristics of large and linear responses over a large field range make these materials attractive as magnetic field sensors, especially in the pulsed magnets where large fields is produced but accurate calibration is a challenge.

\begin{figure}[htb]
\centering
\includegraphics[width=8.5cm]{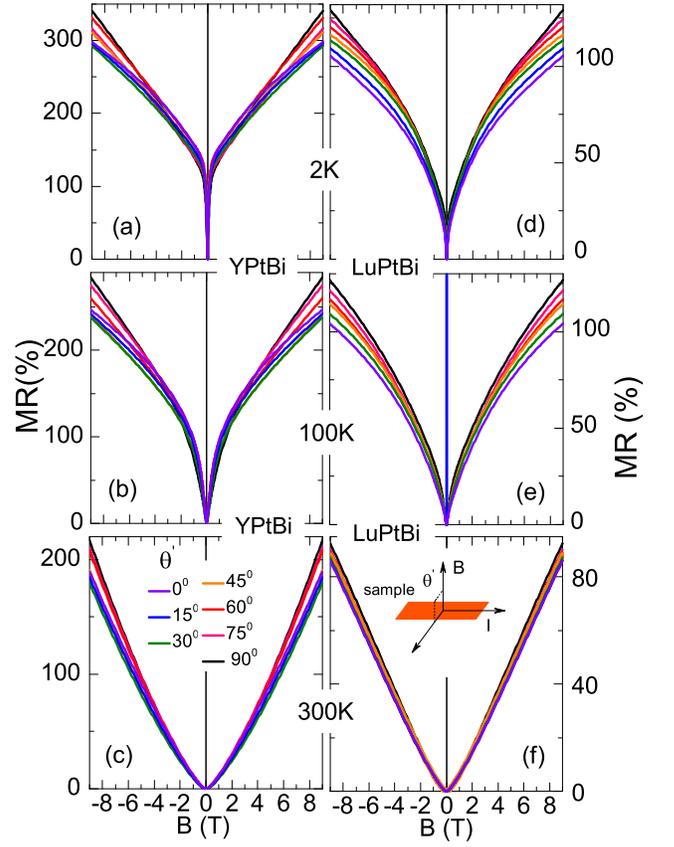}
\caption{(Color online) Isothermal magnetoresistance in varying ($\theta^{'}$) of YPtBi and LuPtBi at (a) (d) 2~K, (b) (e) 100~K , and (c) (f) 300~K. In this case applied magnetic field always perpendicular to the current. Inset of (f) shows sample rotation geometry with field.}
\label{fig:MR_Axis}
\end{figure}

\begin{figure}[htb]
\centering
\includegraphics[width=8.5cm]{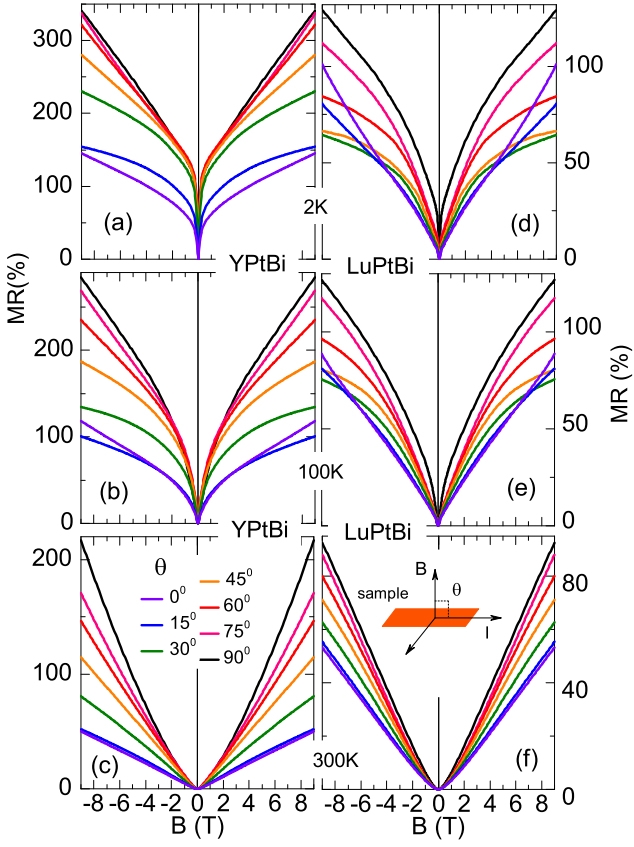}
\caption{(Color online) Isothermal magnetoresistance in varying ($\theta$) of YPtBi and LuPtBi at (a) (d) 2~K, (b) (e) 100~K , and (c) (f) 300~K, where $\theta=0$ (field $||$current) and $\theta=90$ (field $\bot$ current). Inset of (f) shows sample rotation geometry with field.}
\label{fig:MR_Angle}
\end{figure}

Besides the positive and unsaturated MR at high fields, another interesting feature is that materials show fields below 0.5~T, is sharply increase of MR. The sharp increase of the conductivity in low fields, forming a steep cusp, is a well-known signature of the WAL. The WAL  arises in a phase coherent conductor when a destructive interference is formed between two time-reversed electron paths. This destructive interference inhibits elastic backscattering, whereby increases the conductivity. This phase coherency strongly depends on temperatures and fields and the interference is suppressed by an external magnetic field, which leads to a MR in classically weak fields. MR remains a very useful tool for characterizing the phase coherence properties of new materials. Moreover, the topological surface states provide the most favorable host in view of their properties to exhibit WAL. Since this series of Heusler compounds are the topological insulators.~\cite{SC10, XMZ11, CL11} Therefore, it is worthwhile to investigate whether a WAL is also observed in Heusler compounds based TIs. To explore WAL, we measured resistivity in tilting field and resultant MR are shown in Figs.~\ref{fig:MR_Axis} and ~\ref{fig:MR_Angle}. Two types of tilting of field, $\theta$ and $\theta^{'}$ are possible and give the information about surface or bulk dominated transport in WAL regime, respectively. Such rotations are insensitive for ideal 2D conductance. First, MR of Fig~\ref{fig:MR_Axis} was measured by varying $\theta^{'}$ in such way that applied magnetic field always perpendicular to the current and other MR of Fig~\ref{fig:MR_Angle} was measured in varying $\theta$, where $\theta$ is the angle between applied field and current i.e. $\theta=0$ (field $||$current) and $\theta=90$ (field $\bot$ current). The former figure reveals that the MR does not show noticeable change with angle at low fields but slightly changes can be seen at higher fields, while later figure shows that the MR strongly dependent on high fields region. Here, a point should be noted that the steep cusps still appear in both rotations at all angles and temperature below 50~K, that creates a lots of interest to know the its origin.

\begin{figure}[htb]
\centering
\includegraphics[width=8.5cm]{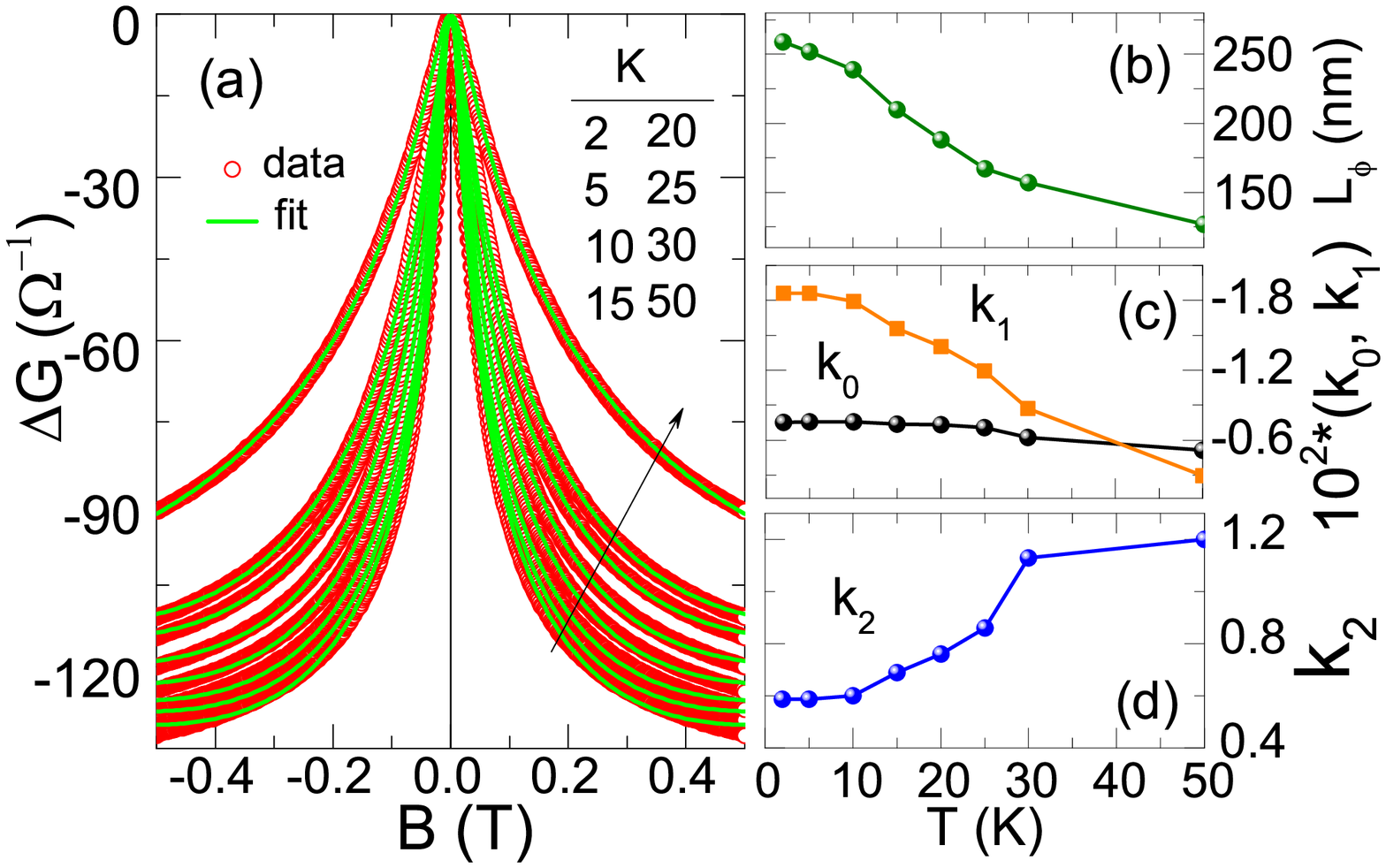}
\caption{(Color online)(a) Isothermal low field transverse conductance of YPtBi at 2, 5 10, 15, 20, 25, 30 and 50~K. Temperature dependence of extracted values of (b) $L_\phi$ (c) $k_0$, $k_1$, and (d) $k_2$.}
\label{fig:YPtBi_fit}
\end{figure}
\begin{figure}[htb]
\centering
\includegraphics[width=8.5cm]{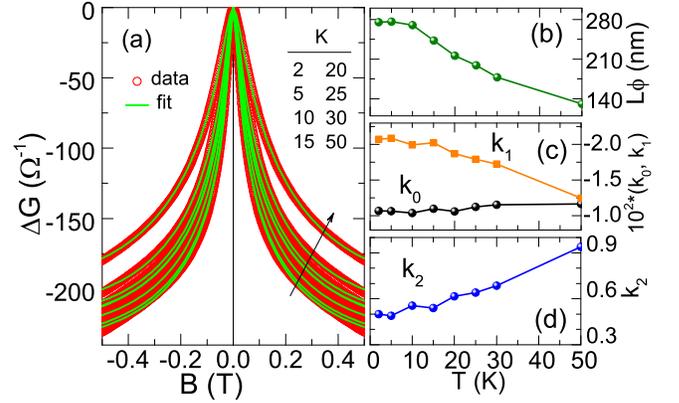}
\caption{(Color online) (a) Isothermal low field transverse conductance of LuPtBi at 2, 5 10, 15, 20, 25, 30 and 50~K. Temperature dependence of extracted values of (b) $L_\phi$ (c) $k_0$, $k_1$, and (d) $k_2$.}
\label{fig:LuPtBi_fit}
\end{figure}
WAL is very commonly observed phenomena in TIs, which reflects the presence of SSs, and we also observed same behavior in bulk crystals of Y(Lu)PtBi TIs. To check the origin of WAL, we explored the magnetoconductivity measurements at low fields and temperatures below 50~K and plotted the conductivity against field, which is shown in Figs.~\ref{fig:YPtBi_fit} and ~\ref{fig:LuPtBi_fit}. This WAL effect is seen only in thin films or nano-flacks of TIs~\cite{ZL12,LB12,ML12} or highly resistive bulk crystal~\cite{CS14}, and originates from strong spin--orbit--coupling resulting spin--momentum locking at the topological SSs.~\cite{KN07} The response of the conductivity with magnetic field in a 2D-system can be quantified by using a well known Hikami--Larkin--Nagaoka (HLN) model. ~\cite{SH80} 

\[\Delta G =G(B)-G(0)\]
\[=\alpha\frac{e^2}{2\pi h}\left[\Psi\left(\frac{1}{2}+\frac{B_\phi}{B}\right)-\ln\left(\frac{B_\phi}{B}\right)\right]\]
\[+\alpha\frac{e^2}{\pi h}\left[\Psi\left(\frac{1}{2}+\frac{B_{so}+B_{e}}{B}\right)-\ln\left(\frac{B_{so}+B_{e}}{B}\right)\right]\]
\[+\alpha\frac{3e^2}{2\pi h}\left[\Psi\left(\frac{1}{2}+\frac{(4/3)B_{so}+B_{e}}{B}\right)-\ln\left(\frac{(4/3)B_{so}+B_{e}}{B}\right)\right]~~(1)\]

where $B_i$ are the characteristic fields of each respective scattering channel ($i=\phi,so,e$) given by $Bi=\hbar/(4eL^2_i)$, $\Psi$ is the digamma function, $L_\phi$ is the phase coherence length, $L_{so}$ is the spin-orbit scattering length, and $L_e$ is the elastic scattering length. At low temperatures and small fields, $L_\phi$ > $L_{so}, L_e$; therefore,  phase coherence scattering dominates  over existing scattering.~\cite{BAA13} Here, $\alpha$ indicates  the type of localization and is equal to $-1$ for WAL in a perfect 2D system.   At high fields, the $L_{so}$ and $L_e$ lengths become prominent and yield characteristic fields of the order of several tesla. The latter two terms containing spin-orbit scattering and elastic scattering can easily be approximated  into a general $B^{k_2}$ term. Therefore, this approximation leads to
\[\Delta G=k_0\left[\Psi\left(\frac{1}{2}+\frac{B_\phi}{B})\right)-\ln\left(\frac{B_\phi}{B}\right)\right]-k_1 B^{k_2}~~~~(2)\]

where, $k_0$($=\alpha\frac{e^2}{2\pi h}$), $k_1$ and $k_2$ are constant and depend on the magnitude and nature of the conductivity of materials. For $k_2=2$, this equation has already been applied earlier in other compounds owing highly bulk resistivity.~\cite{CS14,BAA13} Keeping in mind the behavior of observed magneto-conductivity, it is very interesting to find the temperature dependence values of these constants, which have been used in equation (2). The low fields conductance data are  very well fitted by this equation and are shown by green solid line in Figs.~\ref{fig:YPtBi_fit} (a) and ~\ref{fig:LuPtBi_fit} (a). We obtained the temperature dependence values of fitting parameters $k_0$, $k_1$, $k_2$, and $B_\phi$ and hence $L_\phi$, which are shown in Figs.~\ref{fig:YPtBi_fit} (b), (c), (d) and ~\ref{fig:LuPtBi_fit} (b), (c), (d). The constant $k_0$ is directly connected to $\alpha$, which relates the transport through 2D WAL. The values of $k_0$ at 2~K is -75(-107) for Y(Lu)PtBi and hence the values of $\alpha$ come out to be ordered of $-10^6$, which is irrelevant due to contribution from the bulk and side wall conduction~\cite{CL11,VES14}. More importantly, it is nearly temperature independent, that indicates the constant source of transport, which is either by topological SSs or by the bulk. At this stage, we cannot distinguish it. On the other hand, the value of $L_\phi$ is 245(280)~nm at 2~K for Y(Lu)PtBi, which continuously decreases with increasing temperature, that revels the weakening of WAL at high temperatures. Since, the second term of the equation (2) dominates over the first term in high fields, therefore, it might also be important to draw the conclusion about the value of the constants $k_1$ and $k_2$ at high fields. The value of $k_2$ start from 0.5 at 2~K and reach nearly 1.0 at 50~K, which hint toward bulk localization, where conductivity varies $B^{0.5}$ in high field.~\cite{AK80}. Due to the bulk nature of the samples, that contributes in the transport indicating the origin of WAL is not only because of 2D surface of TIs but it is also due to 3D bulk.~\cite{CS14,HZL11}. Photoemission spectra show that there are many bands cross the Fermi surface including surface bands in this series of compounds, which together contribute in the conductions.~\cite{CL11}  As temperature increases, values of $k_2$ continuously increase for both compounds and become 1.0 i.e. conductivity linearly varies with field, which is another point of interest.

The observed MR can be divided into two regions on field and temperature scales: first, MR at field below 1~T and temperature below 50~K shows strong WAL and it turns into linear above 1~T. Second, MR at temperature above 50~K shows linear behavior with fields. From the above discussion, it is very much clear that the first region of MR governs by WAL while the second part is linear and intriguing to find its origin. Apart from the introduction, there are many different compounds, which are well known for their linear MR. In this group, TIs have recently been added.~\cite{XW12,MK09,AH01,RX97} As discussed in the introduction, Quantum~\cite{AAA98} and classical~\cite{ML02,MMP03}, which are the two well known theories to explain the origin of linear MR. Quantum theory, specially, explain the origin of MR in semi-metal having narrow gap and given as

\[\rho_{xx}=\frac{1}{2\pi}\left(\frac{e^2}{\epsilon_\infty\upsilon}\right)^2 ln\epsilon_\infty\frac{N_i}{ecn_0}B\]

where $n_0$ is the electrons density, $N_i$ is the density of scattering centers, $\epsilon_\infty$ is the dielectric constant at high frequency, and $\upsilon$ is band velocity, which is constant for a band of linear-k dispersion. The key point of quantum MR is that it should be nonsaturating, positive, linear down to low fields and, more interestingly, temperature independent. MR of our semi-metal compounds also shows same trends and more importantly, it is also almost temperature independence up to 50~K, which indicates toward the origin of quantum MR. Indeed the condition for only one Landau level to participate in transport is $n<(eB/c\hbar)^{3/2}$. If we put the $n=6.0\times 10^{18}$~cm$^{-3}$, which is obtained from the Hall measurements, we get B$ > $65~T that is very high to generate the linear MR. The $\beta$-Ag$_{2+\delta}$Te is a well known compound to display linear MR in the large range of temperatures 4.5-300~K and fields 10$^{-3}$-5.5~T.~\cite{RX97}, which has recently been found a TI.~\cite{WZ11,SL12} Formally, it was assumed that the linear energy spectrum may come from the strong disorder, which is introduced by Ag atoms. But now, scenario has changed ever since $\beta$-Ag$_{2+\delta}$Te has been predicted as a TI and it believes that the linear MR come from the topological surface states.~\cite{SL12} However, Y(Lu)PtBi have been investigated as TIs ~\cite{SC10} and exist the metallic surface states ~\cite{CL11} that is a strong evidence, which indicates the quantum origin of linear MR of Y(Lu)PtBi. However, mobility is the origin of the linear MR in this type of compounds that raise the questions about the origin of quantum MR.~\cite{WW13}

\begin{center}
\textbf{IV. CONCLUSIONS}
\end{center}

In conclusion, our grown single crystals of Y(Lu)PtBi are semi-metallic and exhibit high mobility. Temperature below 50~K, both compounds transfer into a quantum transport regime and show weak anti-localization in presence of the strong spin-orbit coupling, which is crucial for a material to be topological insulator. Low field magneto-conductivity behavior clearly show weak anti-localization indicating the electrical conduction through the topological surface states. Besides these, compounds show linear unsaturated magnetorsistance up-to 60~T, which is originated from the mobility. However, the properties like surface states, linear dispersion of spectrum, and low charge carrier of the present compounds, the quantum origin of the linear magnetoresistance cannot simply be ignored. Though, transport through the purely surface states in these materials remains a challenging research work and is necessary to improve for realization of Majorana fermions.

\begin{acknowledgments}

Financial support by the Deutsche Forschungsgemeinschaft (DfG, German Research Foundation) within the priority program SPP1666 ``Topological insulators'' is gratefully
acknowledged..

\end{acknowledgments}



\end{document}